\documentclass[12pt]{iopart}
\usepackage{iopams}
\begin{document}
\title[Stability of differential equations associated with  maps]
{Stability of differential equations associated with a class of 
one dimensional maps}
\author{M. C. Valsakumar$^1$\footnote[3]{To
whom correspondence should be addressed (valsa@igcar.ernet.in)}
A. Rajan Nambiar$^2$ and P Rameshan$^3$}
\address{$^1$Materials Science Division, IGCAR, Kalpakkam, India, 
Pin 603102} 
\address{$^2$Dept. of Physics, Govt. Arts \& Science College, Kozhikode, 
India, Pin 673018}
\address{$^3$Department of Physics, University of Calicut, Calicut 
University PO, India, Pin 673635}
\begin{abstract}
Discrete time evolution of one-dimensional maps is embedded in 
continuous time by truncating the Taylor series expansion of the time 
evolution operator to a finite order N. Truncations with N $>$ 4 leads 
to unconditional instability.  Generalization of the truncated models 
with N = 3 and 4 shows dynamical behaviour characteristic of systems 
with a riddled parameter space
\end{abstract}
\date{\today}
\maketitle
\section{Introduction}

Time development of systems evolving continuously in time is modelled 
with differential equations, whereas difference equations (maps) are 
used to describe systems that evolve in discrete time. Here we try 
to study connection, if any, between solutions of discrete maps and 
a set of ordinary differential equations that are {\it derived} from 
the map using an {\it ansatz} described below. Basically, we truncate 
the Taylor series expansion of the time evolution operator 
corresponding to the discrete map at a finite order $N$ to get an 
ordinary differential equation (ODE) of order $N$. Contrary to the 
expectation that the solution of such an ODE should converge to that 
of the discrete map (when t = an integer) in the limit $N$ $\to$ 
$\infty$ we find surprising results. In particular, all these ODEs  
with $N$ $\ge$ 5 are seen to be unconditionally unstable for almost all 
initial conditions regardless of the details of the underlying map. 
The nature of the solutions for truncation at $N$ = 3 and 4 do depend 
on the details of the map. The $3^{rd}$ and $4^{th}$ order ODEs 
obtained for the logistic map show period doubling bifurcations 
leading to chaos, reverse bifurcations, instability in some regions 
{\it etc.} Unlike the logistic map, these equations show  
sudden switching from regular to chaos or to unstable behaviour 
by a slight variation in the parameter, which is an indication of 
structural instability that was discussed in the context 
of systems with a riddled parameter space (Cazelles 2001, 
Kapitaniak {\it et. al.} 2000, Kim and Lim 2001, Krawiecki and 
Matyjaskiewicz 2001, Lai and Winslow 1994, Lai 2000, Lai and 
Andrade 2001, Madvinsky {\it et. al.} 2001, Maistrenko {\it et. al.} 
1999, Woltering and Markus 2000, Yang 2000, Yang 2001).

\section{Ordinary differential equations corresponding to 
discrete maps}

Consider a general 1-D map 
\begin{equation}
x_{n+1} = f(x_n). 
\label{eqmap}
\end{equation}	    
This discrete time evolution can be embedded in continuous time $t$ by 
considering the equation
\begin{equation} 
\hat{T} x(t) = f(x(t))
\label{eqcont}
\end{equation}	   
where $\hat{T}$ is the unit time evolution operator
\begin{equation} 
\hat{T} = \exp{(\frac{d}{dt})}
\label{eqtimeevol}
\end{equation}	   
We consider the stability of the sequence of equations
\begin{equation} 
\sum_{j=0}^{N}\left(\frac{1}{j!}\right)\left(\frac{d^j}{dt^j}\right) 
x(t) = f(x(t))
\label{eqtrunc}
\end{equation}
obtained by truncating the formal power series expansion of the time 
evolution operator $\hat{T}$ to order $N$.

\section{Stability analysis}
As mentioned in the introduction, one of the objectives of this paper 
is to examine the nature of solutions of the ordinary differential 
equations (\ref{eqtrunc}) for various values of $N$. For any map 
$f(x)$ that is continuous and differentiable, we prove that the 
solutions of \ref{eqtrunc} are linearly unstable for almost all 
initial conditions if $N$ $\ge$ 5. 

\medskip
Consider the stability of solutions of Eq. (\ref{eqtrunc}) about an 
arbitrary reference point (not necessarily a fixed point) $x_{\star}$. 
Writing $x(t)$ = $x_{\star}$ + $\delta x(t)$, one gets
\begin{equation} 
\sum_{j=0}^{N}\left(\frac{1}{j!}\right)\left(\frac{d^j}{dt^j}\right) 
\delta x(t) + \alpha(x_{\star})\delta x(t) = \beta(x_{\star})
\label{eqstabi}
\end{equation}
to first order in $\delta x(t)$. In Eq. (\ref{eqstabi}), 
$\alpha(x_{\star}$ = $1 - \frac{d}{dx}f(x)_{\vert_{x=x_{\star}}}$ and 
$\beta(x_{\star})$ = $f(x_{\star}) -x_{\star}$. Clearly, 
$\beta(x_{\star})$ = 0 if $x_{\star}$ is a fixed point of the map. 
It is easy to show (see appendix A for details) that the stability 
property of the linear inhomogeneous equation (\ref{eqstabi}) is the 
same as that of the homogeneous part
\begin{equation} 
\sum_{j=0}^{N}\left(\frac{1}{j!}\right)\left(\frac{d^j}{dt^j}\right) 
\delta x(t) + \alpha(x_{\star})\delta x(t) = 0.
\label{eqstab}
\end{equation}
Since the above equation is linear we look for a solution of the form
\begin{equation} 
\delta x(t) = c \exp{(\mu t)}
\label{eqexp}
\end{equation}
to get 
\begin{equation} 
\sum_{j=0}^{N}\left(\frac{1}{j!}\right) \mu^j + \alpha(x_{\star}) =0.
\label{eqlinear}
\end{equation}
We now show that there will exist at least one $\mu$ with positive 
real part irrespective of the value of $\alpha(x_{\star})$, if $N$ $\ge$ 5. 
This would imply that the solution of the equation (\ref{eqtrunc}) 
is unstable for any finite $N$ $\ge$ 5 for almost any initial 
condition, regardless of the specific functional form of $f(x)$. 

Now, Eq. (\ref{eqlinear}) is of the form
\begin{equation} 
a_0 \mu^N + a_1 \mu^{N-1} + ........ + a_{N-1}\mu + a_N =0,
\label{eqrouth1}
\end{equation}
with $a_j$ = $\frac{1}{(N-j)!}$ for $0$ $\le$ j $\le$ $(N-1)$  and 
$a_N$ = $\alpha(x_{\star})$.

\medskip
Nature of zeros of Eq. (\ref{eqrouth1}) can be examined by using 
the well known Routh-Hurwitz theorem (Korn and Korn 1961). Define	 
\begin{eqnarray}
\nonumber
U_0 = a_0, \ \ U_1 = a_1, \ \ 
U_2 = \left\vert
\begin{array}{cc}
a_1 & a_0\\a_3 & a_2
\end{array}
\right\vert,\\
U_3 = \left\vert
\begin{array}{ccc}
a_1 & a_0 & 0\\a_3 & a_2 & a_1\\a_5 & a_4 & a_3
\end{array}
\right\vert,
U_4 = \left\vert
\begin{array}{cccc}
a_1 & a_0 & 0 & 0\\a_3 & a_2 & a_1 & 0\\a_5 & a_4 & a_3 & a_2\\a_7 & a_6 & a_5 & a_4
\end{array}
\right\vert, {\ \ \rm etc.}
\label{eqrouth2}
\end{eqnarray}
Then Routh-Hurwitz theorem states that the number of roots of Eq. 
(\ref{eqrouth1}) with positive real parts is equal to the number 
of sign changes in the sequence $\lbrace U_j \rbrace$. If at least 
one sign change occurs in the sequence $\lbrace U_j \rbrace$, then 
Eq. (\ref{eqrouth1}) has at least one root ($\mu$) with positive real 
part and hence the solution of Eq. (\ref{eqtrunc}) is unstable in the 
neighbourhood of the reference point $x_{\star}$. If this happen for 
almost any $x_{\star}$, then the solution is unstable for any initial 
condition, globally. 

\medskip
We now calculate the sequence $\lbrace U_j \rbrace$ when $N$ $>$ 5. 
Substituting for $a_j$ in Eq. (\ref{eqrouth2}), and on simplifying, 
we get
\begin{equation}
U_0 = \frac{1}{N!}, \ \ U_1 = \frac{1}{(N-1)!}, \ \ 
U_2 = \frac{2}{N!(N-2)!},
\end{equation}
which are positive for all positive $N$ and  
\begin{equation}
U_3 = -\frac{2(N-5)}{N!(N-1)!(N-3)!} 
\end{equation}
which is negative for all $N$ $>$ 5. Since there is at least one sign 
change in the sequence $\lbrace U_j \rbrace$, it follows that 
the solutions of Eq. {\ref{eqtrunc}) are linearly unstable for 
$N$ $>$ 5 irrespective of the details of the mapping function $f(x)$, 
so long as it is continuous and differentiable. Since the reference 
point $x_{\star}$ does not explicitly appear in arriving at this 
conclusion, it is evident that the solutions of Eq. (\ref{eqtrunc}) are 
unstable for almost any initial condition for $N$ $>$ 5.

We now turn to the case $N$ = 5. Explicit calculation shows that
\begin{equation}
\fl U_0 = \frac{1}{5!}, \ U_1 = \frac{1}{4!}, \ U_2 = \frac{2}{5! 3!}, 
\  U_3 = \frac{\alpha(x_{\star}) -1}{5! 4!}  \ {\rm and \ } 
U_4 = -\frac{(\alpha(x_{\star}) -5/3)^2 + 20/9}{5!^2}
\label{eqrouth3}
\end{equation}
From Eq. (\ref{eqrouth3}), it is clear that there is a sign change at 
$U_3$ and no further sign change at $U_4$, if $\alpha(x_{\star})$ $<$ 1. On the 
otherhand, if $\alpha(x_{\star})$ $>$ 1, $U_3$ $>$ 0 and $U_4$ $<$ 0. Thus there 
is at least one sign change in the sequence $\lbrace U_j \rbrace$, 
regardless of the details of the mapping function $f(x)$ for $N$ = 4, 
as well. No such general conclusions can be drawn for the cases $N$ = 1, 2, 3 
and 4. 

\medskip
We now turn to the specific case of truncation of chaotic maps. Since a 
minimum of three dimensions is required for occurrence of chaos in 
ordinary differential equations, truncations to order $N$ = 1 and 2 are 
uninteresting from this perspective. We also now know that the 
truncations to order $N$ $\ge$ 5 lead to unconditional instability. 
Therefore, in what follows,  we concentrate on truncations with 
$N$ = 3 and 4. It is found (Rajan Nambiar 2003) that the solutions 
of these equations show many features which are not shared by the 
solutions of the original discrete map. For concreteness, we present 
the results for the logistic map $f(x)$ = $px(1-x)$, $x$ $\epsilon$ 
$[0, 1]$.

\section{Truncation at N=3 and 4}

For this case, Eq. (\ref{eqtrunc}) becomes
\begin{equation}
\frac{d^3x}{dt^3} + 3 \frac{d^2x}{dt^2} + 6 \frac{dx}{dt} + 6 (x-f(x)) = 0
\label{eqtrunc3}
\end{equation}
Linear stability analysis of the above equation, with $f(x)$ = 
$p x (1-x)$ shows that both fixed points (0  and $1-1/p$) are stable 
for $p$ $<$ 4. Thus the truncation at $N$ = 3 leads to regular 
behaviour for all $p$ $\epsilon$ $[0, 4]$, the range of $p$ for which 
the logistic map is map of the unit interval to itself. This has to 
be contrasted with the occurrence of chaos in the logistic map when  
$p$ $\epsilon$ $[3.66..., 4]$. However, the system of equations 
(\ref{eqtrunc3}) do show chaotic behaviour when $p$ $>$ 4.

\medskip
Using the scaled variables $X$= $(2p/9)x$ and $\tau$ =$t/3$, Eq. 
(\ref{eqtrunc3}) can be rewritten as 
\begin{equation}
\frac{d^3X}{d\tau^3} + \frac{d^2X}{d\tau^2} + \nu \frac{dX}{d\tau} + 
-\lambda X + X^2 = 0
\label{eqtrunc3a}
\end{equation}
where $\nu$ = 2/3 and $\lambda$ = $2(p-1)/9$. In fact, the generalization 
of Eq. (\ref{eqtrunc3a}) obtained by allowing $\nu$ to take arbitrary 
values is equivalent to the equations studied by Coulett {\it et. al.} 
(Coulett {\it et. al.} 1979) and Arneodo {\it et. al.} (Arneodo 
{\it et. al.} 1985) which exhibited  striking features in its solution. 
The model shows regular, chaotic and or unstable behaviour for certain 
choices of the parameters $\nu$ and $\lambda$. For example, for a fixed $\nu$, 
as $\lambda$ is increased, we observe sequences of either finite or infinite number 
of period doubling bifurcations and reverse bifurcations with unstable 
regions in between. The bifurcation diagram is substantially different 
for a neighbouring $\nu$. Thus the significant aspect of this equation is 
that the nature of solution changes drastically for very small changes in 
the values of the parameters. It is as if the parameter space is riddled. 
The solution of ODE obtained by truncating at $N = 4$ also show similar features.

\medskip
\section{Conclusions}
The present work shows that the nature of solutions of the ordinary 
differential equations, obtained by truncating the power series of 
the time evolution operator corresponding to discrete maps, are very 
different from those of the discrete maps. In particular, the 
solutions with $N$ $\ge$ 5 are unconditionally unstable regardless 
of the details of the map. For the specific case of the logistic map, 
truncations at $N$ = 3 and 4 lead to solutions characteristic of systems 
with a riddled parameter space.
\section{Acknowledgements}
Rajan Nambiar is grateful to U.G.C. and Govt. of Kerala for granting 
him leave under FIP for this study.

\section*{References}
\begin{harvard}
\item[]Arneodo A, Coulett P H, Spiegel E A and Tresser C 1985  
{\it Physica} {\bf D 14} 327.
\item[]Cazelles B 2001 {\it Phys. Rev.} {\bf E 64} 032901.
\item[]Coulett P, Tresser C  and Arneodo A 1979 {\it Phys. Lett.} 
{\bf A 72} 268.
\item[]Kapitaniak T, Maistrenko Y and Popovych S 2000  
{\it Phys. Rev.} {\bf E 62} 1972.
\item[]Kim S T and Lim W 2001  {\it Phys. Rev.} {\bf E 63} 026217.
\item[]Korn G A and Korn TM 1961 {\it Mathematical Handbook for 
Scientists and Engineers} McGraw-Hill Book Co., New York.
\item[]Krawiecki A and Matyjaskiewicz S 2001 {\it  Phys. Rev.} 
{\bf E 64} 036216.
\item[]Lai Y C and Winslow R L 1994 {\it Phys. Rev. Lett.} 
{\bf 72} 1640.
\item[]Lai Y C 2000 {\it Phys. Rev.} {\bf E 62} R4505.
\item[]Lai Y C and Andrade V 2001 {\it Phys. Rev.} {\bf E 64} 056228.
\item[]Madvinsky A B  {\it et. al.} 2001 {\it Phys. Rev.} 
{\bf E 64} 021915.
\item[]Maistrenko Yu L, Maistrenko V L, O. Popovych and  
Mosekilde E 1999 {\it Phys. Rev.} {\bf E 60} 2817.
\item[]Rajan Nambiar A 2003 {\it Ph.D. Thesis, University of Calicut}
\item[]Woltering M and Markus M 2000 {\it Phys. Rev. Lett.} 
{\bf 84} 630.
\item[]Yang H L 2000 {\it Phys. Rev.} {\bf E 62} R4509.
\item[]Yang H L 2001  {\it Phys. Rev.} {\bf E 63}  036208.
\end{harvard}
\appendix
\section{Solution of the inhomogeneous equation (\ref{eqstabi})}
Let us define an $N$- dimensional column vector \textbf{$\xi$}
\begin{equation}
\fl \textbf{$\xi$} =\left(\xi_1, ... , \xi_N\right)^T, \ \ {\rm with \ \ } 
\xi_1=\delta x(t), \ \ \xi_{j+1} = \frac{d}{dt}\xi_j, \ \ 
j = 1,...., (N-1)
\label{A1}
\end{equation}
Equation (\ref{eqstabi}) can then be rewritten as 
\begin{equation}
\frac{d}{dt}\textbf{$\xi$}(t) = \textbf{M} \textbf{$\xi$}(t) + \textbf{v}
\label{A2}
\end{equation}
where 
\begin{equation}
\fl \textbf{M} = \left(\begin{array}{ccccccc}
0    & 1    & 0    & 0    & .... & 0\\
0    & 0    & 1    & 0    & .... & 0\\
.... & .... & .... & .... & .... & .... \\
0    & 0    & 0    & 0    & .... & 1\\
-\frac{a_N}{a_0}&-\frac{a_{N-1}}{a_0}&-\frac{a_{N-2}}{a_0}&
-\frac{a_{N-3}}{a_0}& .... & -\frac{a_1}{a_0}
\end{array}\right), \ \ {\rm and \ \ } \textbf{v} = 
\left(\begin{array}{c}
0\\
0\\
....\\
0\\
\beta(x_{\star})
\end{array}\right)
\label{A3}
\end{equation}
The formal solution of Eq. (\ref{A2}) is given by
\begin{equation}
\textbf{$\xi$}(t) = \exp{[\textbf{M}t]} \textbf{$\xi_0$} + 
\int_0^t dt' \exp{[\textbf{M}(t-t')]} \textbf{v}(t')
\label{A3a}
\end{equation}
Using the fact that \textbf{v} is independent of time and redefining 
the variable of integration, we get
\begin{equation}
\textbf{$\xi$}(t) = \exp{[\textbf{M}t]} \textbf{$\xi_0$} + 
\left(\int_0^t dt' \exp{[\textbf{M} t']}\right) \textbf{v}
\label{A4}
\end{equation}
Let \textbf{S} be the similarity transformation matrix that 
diagonalises \textbf{M} so that
\begin{equation}
\textbf{S}^{-1} \textbf{M} \textbf{S} = \textbf{D} = Diag(\mu_1, ....., \mu_N)
\label{A5}
\end{equation}
It then follows that 
\begin{equation}
\fl \textbf{$\xi$}(t) =\textbf{S} Diag(e^{\mu_1 t}, ....... , e^{\mu_N t}) 
\textbf{$\xi_0$} + \textbf{S} Diag\left(\frac{e^{\mu_1 t}-1}{\mu_1}, 
........, \frac{e^{\mu_N t}-1}{\mu_N}\right) \textbf{S}^{-1} \textbf{v}
\label{A6}
\end{equation}
The first term on the right hand side of Eq. (\ref{A6}) corresponds to 
the solution of the homogeneous part (see Eq. (\ref{eqstab})). It is clear 
that if at least one of the $\lbrace \mu_j\rbrace$ has a positive real 
part so that the solution of the homogeneous equation is unstable, so 
is the solution of the inhomogeneous equation.
\end{document}